\documentclass[showkeys,showpacs,superscriptaddress,nofootinbib]{revtex4-2}
\usepackage{amsmath}
\usepackage{amssymb}
\usepackage{graphicx}
\usepackage{bm}
\usepackage{float}
\usepackage{ulem,xcolor}
\usepackage{color,soul}

\begin{document}

\title{
	QCD effects in non-QCD theories
 }

\author{
Vladimir Dzhunushaliev
}
\email{v.dzhunushaliev@gmail.com}


\affiliation{
Department of Theoretical and Nuclear Physics,  Al-Farabi Kazakh National University, Almaty 050040, Kazakhstan
}

\affiliation{
	Institute of Experimental and Theoretical Physics,  Al-Farabi Kazakh National University, Almaty 050040, Kazakhstan
}

\affiliation{
Academician J.~Jeenbaev Institute of Physics of the NAS of the Kyrgyz Republic, 265 a, Chui Street, Bishkek 720071, Kyrgyzstan
}

\author{Vladimir Folomeev}
\email{vfolomeev@mail.ru}

\affiliation{
	Institute of Experimental and Theoretical Physics,  Al-Farabi Kazakh National University, Almaty 050040, Kazakhstan
}

\affiliation{
	Academician J.~Jeenbaev Institute of Physics of the NAS of the Kyrgyz Republic, 265 a, Chui Street, Bishkek 720071, Kyrgyzstan
}


\date{\today}

\begin{abstract}
It is shown that, in some non-QCD theories, there are effects shared by QCD: (i)~in SU(2) Yang-Mills theory containing a nonlinear spinor field, there is a mass gap; (ii)~in SU(3) Proca-Higgs theory, there are flux tube solutions with a longitudinal electric field required for producing a force binding quarks; (iii)~in non-Abelian Proca-Higgs theories, there exist flux tube solutions with a momentum directed along the tube axis and
particlelike solutions with a nonvanishing total angular momentum created by crossed color electric and magnetic fields; in QCD, such configurations may contribute to the proton spin. We discuss the conjecture  that such non-QCD theories might be a consequence of approximate solution of an infinite set of Dyson-Schwinger equations describing the procedure of nonperturbative quantization. The phenomenon of dimensional transmutation for nonperturbative quantization and the analogy between nonperturbative quantization and turbulence modeling are also discussed.
\end{abstract}



\date{\today}

\maketitle

\section{Introduction}

In quantum chromodynamics (QCD), there are many effects caused by the strong nonlinearity of QCD. This means that such effects must be described using nonperturbative quantization of strongly nonlinear non-Abelian Yang-Mills fields. These effects are, for instance,
(i)~the appearance of a mass gap,
(ii)~the existence of tubes connecting quarks and filled with a longitudinal electric field,
(iii)~the presence of a contribution to the proton spin coming from gluon fields, etc. The presence of the mass gap indicates that, for quantized Yang-Mills fields, there exists a lower limitation on the mass of particles.
The existence of a mass gap in QCD is not an exceptional phenomenon. For example, in the BCS theory of superconductivity, there is an energy gap whose existence
is caused by the energy profit for a pair of electrons creating a Cooper pair (evaluating the energy gap for superconductors can be found in textbooks, see, e.g., Ref.~\cite{tinkham}).
In the Nambu-Jona-Lazinio model, there appears the so-called equation for a mass gap according to which the mass spectrum of mesons starts not from zero but from
some nonzero value which is called a mass gap (see, e.g., the review~\cite{Klevansky:1992qe}). In turn,
the existence of the aforementioned tubes is related to the confinement phenomenon (for a detail review of the confinement problem, see, e.g., Ref.~\cite{Greensite:2011zz}):
such a tube provides so strong attraction between quarks that they are never found in isolation. It may be noted here that, in the MIT bag model, there is a string approximation for a
flux tube connecting quarks and containing a longitudinal color electric field (for a review of this model, see Ref.~\cite{Hasenfratz:1977dt}).

For its part, it was experimentally observed that there exist contributions to the proton spin coming from quark spins, their orbital angular momenta, as well as gluon fields~\cite{Ashman:1987hv,Ashman:1989ig}. Nearly all these phenomena are described using lattice calculations (for a detail review on lattice calculations see  Ref.~\cite{Ratti:2018ksb}). However, in a certain sense, lattice calculations are similar to an experiment: they give a certain result that one would like to explain using some analytic calculations.

In the present paper we wish to discuss the possibility that, in some non-QCD theories, there may exist effects shared by QCD (here, by QCD, we mean SU(3) gauge theory with quarks). The appearance of such QCD effects in non-QCD theories enables one to assume that such non-QCD theories may have some relevance to QCD. For example, it is possible that such non-QCD theories may serve as some {\it approximate} way to describe nonperturbative effects in QCD. This would imply that, in nonperturbatively quantized QCD, some approximations leading to non-QCD theories are possible.

As concrete examples, we will consider here, firstly, the possibility of the appearance of a mass gap in SU(2) Yang-Mills theory where color fields are sourced by a nonlinear spinor field (Sec.~\ref{mass_gap}). Second, in Sec.~\ref{Proca_theories}, we will demonstrate the possibility of obtaining various  flux tube solutions (within SU(3) Proca-Higgs theory) and particlelike solutions with a nonzero total angular momentum (within SU(2) Proca-Higgs theory).  Such configurations may be used in modeling the confinement phenomenon and contribute to the proton spin.

\section{Mass gap}
\label{mass_gap}

In this section we demonstrate that the energy spectrum of a ``monopole'' in SU(2) Yang-Mills theory with a source in the form of a nonlinear spinor field has a global minimum corresponding to a mass gap. We have enclosed the word ``monopole'' in quotation marks because the asymptotic behavior of the corresponding solution (sourced by a nonlinear spinor field) differs from that of the  't~Hooft-Polyakov monopole solution (sourced by a triplet of Higgs scalar fields). (Detailed description of the properties of magnetic monopoles can be found, for example, in Ref.~\cite{Shnir:2005vvi}, and for a review on topological methods in QCD
and of the literature in the field see Ref.~\cite{Shuryak:2021vnj}). Also, we will discuss the possibility of the appearance in the theory of a nonlinear spinor field as a consequence of some approximate description of
nonperturbative effects in non-Abelian quantum field theory.

\subsection{Energy spectrum of the ``monopole''}

Here, we closely follow Refs.~\cite{Dzhunushaliev:2020qwf, Dzhunushaliev:2021apa}. The Lagrangian describing a system consisting of a non-Abelian SU(2) field $A^a_\mu$ interacting with nonlinear spinor field $\psi$ can be taken in the form
\begin{equation}
	\mathcal L = - \frac{1}{4} F^a_{\mu \nu} F^{a \mu \nu}
	+ i \hbar c \bar \psi \gamma^\mu D_\mu \psi  -
	m_f c^2 \bar \psi \psi+
	\frac{\Lambda}{2} g \hbar c \left( \bar \psi \psi \right)^2.
\label{1_10}
\end{equation}
Here $m_f$ is the mass of the spinor field;
$
D_\mu = \partial_\mu - i \frac{g}{2} \sigma^a
A^a_\mu
$ is the gauge-covariant derivative, where $g$ is the coupling constant and $\sigma^a$ are the SU(2) generators (the Pauli matrices);
$
F^a_{\mu \nu} = \partial_\mu A^a_\nu - \partial_\nu A^a_\mu +
g \epsilon_{a b c} A^b_\mu A^c_\nu
$ is the field strength tensor for the SU(2) field, where $\epsilon_{a b c}$ (the completely antisymmetric
Levi-Civita symbol) are the SU(2) structure constants;  $\Lambda$ is a constant; $\gamma^\mu$ are the Dirac matrices in the standard representation; $a,b,c=1,2,3$ are color indices and $\mu, \nu = 0, 1, 2, 3$ are spacetime indices.

Using the Lagrangian~\eqref{1_10}, one can find the corresponding field equations
\begin{align}
	D_\nu F^{a \mu \nu} = & \frac{g \hbar c}{2}
	\bar \psi \gamma^\mu \sigma^a \psi ,
\label{1_20}\\
	i \hbar \gamma^\mu D_\mu \psi  - m_f c \psi + \Lambda g \hbar \psi
	\left(
	\bar \psi \psi
	\right) = & 0.
\label{1_25}
\end{align}
In order to find  ``monopole'' solutions and their energy spectrum, we choose the following {\it Ans\"atze} for the Yang-Mills and spinor fields:
\begin{eqnarray}
A^a_t &=& 0, \quad	A^a_i =  \frac{1}{g} \left( 1 - f \right)
	\begin{pmatrix}
		0 & \phantom{-}\sin \varphi &  \sin \theta \cos \theta \cos \varphi \\
		0 & -\cos \varphi &   \sin \theta \cos \theta \sin \varphi \\
		0 & 0 & - \sin^2 \theta
	\end{pmatrix}  \, \text{with} \,\,
	i = r, \theta, \varphi  \text{ (in polar coordinates)},\nonumber
\label{1_30}\\
	\psi^T &=& \frac{e^{-i \frac{E t}{\hbar}}}{g r \sqrt{2}}
	\begin{Bmatrix}
		\begin{pmatrix}
			0 \\ - u \\
		\end{pmatrix},
		\begin{pmatrix}
			u \\ 0 \\
		\end{pmatrix},
		\begin{pmatrix}
			i v \sin \theta e^{- i \varphi} \\ - i v \cos \theta \\
		\end{pmatrix},
		\begin{pmatrix}
			- i v \cos \theta \\ - i v \sin \theta e^{i \varphi} \\
		\end{pmatrix}
	\end{Bmatrix},\nonumber
\label{1_40}
\end{eqnarray}
where $E/\hbar$ is the spinor frequency and the functions $f, u$, and $v$ depend on the radial coordinate $r$ only. Substituting these {\it Ans\"atze} in Eqs.~\eqref{1_20} and \eqref{1_25}, one can find the corresponding equations for the functions $f, u$, and $v$ (for details see Ref.~\cite{Dzhunushaliev:2020qwf}).

\begin{figure}[H]
\centering
\includegraphics[width=.5\linewidth]{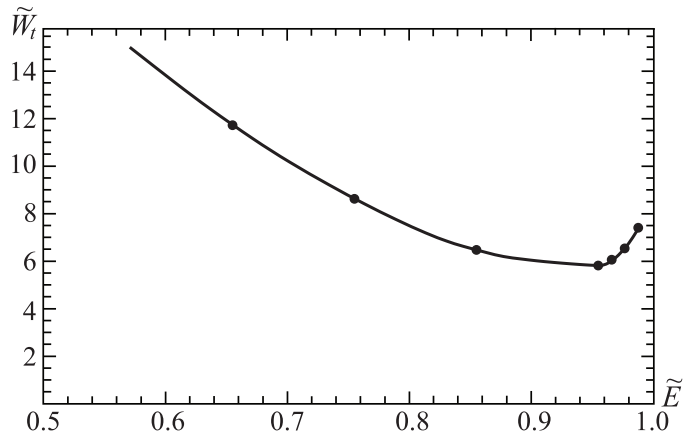}
\caption{A sketch of the energy spectrum of the total energy from Eq.~\eqref{1_70} as a function of the spinor frequency $\tilde E$
(for details see Ref.~\cite{Dzhunushaliev:2020qwf}).
}
\label{energy_spectrum}
\end{figure}

In turn, the total dimensionless energy density of the ``monopole'' under consideration is
\begin{equation}
	\tilde\epsilon =	 \frac{1}{\tilde g^2_\Lambda}
	\left[
	\frac{{f'}^2}{ x^2} +
	\frac{\left( f^2 - 1 \right)^2}{2 x^4}
	\right] +
	\left[
	\tilde E \frac{\tilde u^2 + \tilde v^2}{x^2} +
	\frac{\left(\tilde u^2 - \tilde v^2 \right)^2}{2 x^4}
	\right],
	\label{1_60}
\end{equation}
written in terms of the dimensionless variables
$x = r/\lambda_c$,
$
\tilde u=u\sqrt{\Lambda/\lambda_c g},
\tilde v = v\sqrt{\Lambda/\lambda_c g},
\tilde E = \lambda_c E/(\hbar c),
\tilde g^2_{\Lambda} = g \hbar c\lambda_c^2/\Lambda
$, where $\lambda_c= \hbar / (m_f c)$ is the Compton wavelength. The prime denotes differentiation with respect to  $x$. Correspondingly, the total dimensionless energy of the ``monopole'' can be calculated using the formula
\begin{equation}
	\tilde W_t \equiv \frac{\lambda_c g^2}{\tilde g^2_\Lambda} W_t =
	4 \pi
	\int\limits_0^\infty x^2 \tilde \epsilon d x	,
	\label{1_70}
\end{equation}
where the energy density $\tilde \epsilon$ is taken from Eq.~\eqref{1_60}.
The corresponding profile of the total energy is given in  Fig.~\ref{energy_spectrum}.

\subsection{Outline of nonperturbative quantization and possible appearance of a nonlinear spinor field}
\label{NP_quantization}

As we see from Fig.~\ref{energy_spectrum}, the energy of the ``monopole'' supported by the nonlinear spinor field has a minimum which corresponds to a mass gap. In this connection the natural question arises: whether there exists a relation between the mass gap obtained and a mass gap in QCD? In other words,
whether the appearance of the mass gap for such a ``monopole'' solution is coincidental or this is related somehow to nonperturbative quantization of essentially nonlinear non-Abelian Yang-Mills fields? If one starts from the conjecture that such a relation does really exist, it may consist in the fact that the nonlinear spinor field occurs as a consequence of some approximate description of quantum nonpertubative effects in QCD.

This may happen as follows. According to W.~Heisenberg~\cite{heis}, the procedure of nonperturbative quantization
consists in writing out quantum averages of the Yang-Mills and Dirac operator equations. But since these equations contain
Green's functions of different orders, to close the system, one has to write out equations for higher-order Green's functions, and so on, {\it ad infinitum}.
As a result, this leads to an infinite set of Dyson-Schwinger-type equations for all Green's functions.
This has been discussed in detail in Ref.~\cite{Bender:1999ek}, where
 a technique based on truncating the Dyson-Schwinger equations for quantum $\phi^4$ field theories is presented.

In a more rigorous mathematical language, this looks like
\begin{align}
	i \gamma^\mu \left\langle
		\hat \psi_{;\mu} - m \hat \psi
	\right\rangle = & 0 ,
\label{NP_10}\\
	\left\langle \hat F^{a \mu \nu} \right\rangle_{; \nu}
	= & - 4 \pi \left\langle \hat j^{a \mu} \right\rangle ,
\label{NP_20}\\
	\left\langle
		\hat{\bar \psi} \left(
			 i \gamma^\mu \hat \psi_{;\mu} - m \hat \psi
			 \right)
		\right\rangle = & 0 ,
\label{NP_30}\\
\left\langle \hat A_\nu
		\left( 	i \gamma^\mu \hat \psi_{;\mu} - m \hat \psi
		\right)
	\right\rangle = & 0 ,
\label{NP_40}\\
	\left\langle \hat F^{a \mu \nu} \hat \psi \right\rangle_{; \nu}
	= & - 4 \pi \left\langle \hat j^{a \mu} \hat \psi \right\rangle ,
\label{NP_50}\\
	\ldots
\label{NP_60}
\end{align}
Here, the two first equations are the quantum-averaged Dirac and Yang-Mills operator equations, and all subsequent ones are gotten by multiplying by the corresponding operators with subsequent quantum averaging. It must be mentioned that these equations are well known in perturbative quantum field theory, but there they are {\it always} written down using Feynman diagrams, and this leads to {\it a fundamental difference between the Dyson-Schwinger equations in perturbative and nonperturbative regimes.}
The same equations have been obtained in Ref.~\cite{Frasca:2019ysi}, using only another notations for Green's functions. Notice also that, within this approach, it is of course necessary to study the problem of unitary evolution for these equations. 

It is of great importance to note that these equations greatly simplify in the stationary case. Apparently, in this case all Green's functions will be regular, without involving Dirac delta functions, step functions, etc. This situation differs in principle from the case of quantum fields changing with time where one must take into account the finiteness of the velocity of light; this results in the appearance of singular Green's functions.
Such a situation occurs, for example, in perturbative quantum  field theory where Green's functions are singular functions occurring as a result of the fact that in such a case Green's functions are polylinear combinations of propagators. In turn, the propagators are used for a mathematical description of dynamical processes of a motion and of propagation of quanta.

It is evident that such an infinite set of equations can be solved neither analytically nor numerically. Therefore, it is necessary to have some approximate methods of solving Eqs.~\eqref{NP_10}-\eqref{NP_60}.  One of such methods consists in using the procedure that enables one to cut off an infinite set of equations to get a finite one. In this case it is necessary to have a hypothesis about the behavior of higher-order Green's function. In turbulence modeling, this problem is called the closure problem. We will discuss the relation between nonperturbative quantization and turbulence modeling in more detail below.

Consider a possible scenario of the occurrence of nonlinearity of a spinor field. One of the equations in the system~\eqref{NP_10}-\eqref{NP_60} contains a nonlinearity describing the interaction between quarks and the non-Abelian gauge field,
\begin{equation}
	\mathcal L_i \sim \left\langle
		\hat{\bar \psi} \hat{A}^a_\mu \hat \psi
	\right\rangle .
\label{1_b_10}
\end{equation}
If we wish to cut off the infinite set of equations \eqref{NP_10}-\eqref{NP_60} at just the point where one may derive separate equations for the gauge field  $\left\langle A^a_\mu \right\rangle $ and for the spinor field $\left\langle \psi \right\rangle $, we have to have a hypothesis about an approximate description of the interaction~\eqref{1_b_10}. For this purpose, one can suggest different approximations for \eqref{1_b_10}, one of which is
 \begin{equation}
	\left\langle
		\hat{\bar \psi} \gamma^\mu \hat{A}^a_\mu \hat \psi
	\right\rangle =
	\left\langle
		\hat{\bar \psi} \gamma^\mu \left(
			\left\langle \hat{A}^a_\mu \right\rangle + \widehat{\delta A^a_\mu}
		\right) \hat \psi
	\right\rangle
	\approx
	\left\langle \hat{\bar \psi} \gamma^\mu \hat \psi \right\rangle \left\langle A^a_\mu\right\rangle
	+ \lambda
	\begin{cases}
			\left\langle \hat{\bar{\psi}} \hat \psi \right\rangle
			\left\langle \widehat{\delta A^a_\mu} \widehat{\delta A^{a \mu}} \right\rangle ,	\\
			\left\langle \hat{\bar{\psi}} \hat \psi \right\rangle^2 , \\
			\left\langle \hat{\bar{\psi}} \gamma^\mu \hat \psi \right\rangle
			\left\langle \hat{\bar{\psi}} \gamma_\mu \hat \psi \right\rangle ,
	\end{cases}
\label{sp_approximation}
\end{equation}
where the last term on the right-hand side represents possible corrections to the first term needed
for an approximate description of the quantum average given on the left-hand side of this expression;
the closure constant $\lambda$ appears here for matching dimensions of all terms. The physical meaning of such an approximation is that the quantum average of the term describing the interaction of two fields can be approximately represented as a sum of two terms. The first term represents the product of two averages of these fields, and the second one~--  a nonperturbative quantum correction.

It is of great interest that, in this approximation, there occurs the new  dimensional constant $\lambda$. The presence of such a dimensional constant is, perhaps, an example of the fact that there is dimensional transmutation in a {\it nonperturbative} regime of quantization, by analogy with the phenomenon of dimensional transmutation occurring in a {\it perturbative} regime when a dimensional constant appears in a theory where it was initially absent~\cite{Coleman:1973jx}.

\section{Flux tubes and particlelike configurations in Proca theories}
\label{Proca_theories}

In this section, we will show that, in Proca theories, there are such QCD effects:
\begin{itemize}
	\item A tube filled with a longitudinal non-Abelian electric field. Such a tube is a necessary ingredient of the confinement phenomenon, since it provides a constant force between quarks preventing their separation.
	\item A tube filled with crossed color electric and magnetic fields that give a nonzero Poynting vector, and hence a momentum directed along the tube axis.
Three such tubes, binding quarks together in a proton, will contribute correspondingly to the proton spin.
	\item Particlelike solutions possessing a nonvanishing angular momentum created by static crossed non-Abelian electric and magnetic fields.
It is possible that this may have a connection with the nature of the angular momentum of gluon fields contributing to the proton spin.
\end{itemize}

\subsection{Tubes with the flux of chromoelectric field
}
\label{tube_1}

In Ref.~\cite{Dzhunushaliev:2019sxk}, within non-Abelian Proca theory with a scalar field, there were obtained
cylindrically symmetric solutions describing infinite tubes with the flux of a color electric field.
Such tubes can be used for a description of the confinement phenomenon, since they contain a longitudinal electric field producing a constant attractive force between quarks and preventing their separation.

The Lagrangian describing such a system consisting of a non-Abelian SU(3) Proca field $A^a_\mu$ interacting with nonlinear scalar field $\phi$ can be taken in the form (in this section we work in units such that $c=\hbar=1$)
\begin{equation}
	\mathcal L =  - \frac{1}{4} F^a_{\mu \nu} F^{a \mu \nu} -
	\frac{\left( \mu^2 \right)^{a b, \mu}_{\phantom{a b,}\nu}}{2}
	A^a_\mu A^{b \nu} +
	\frac{1}{2} \partial_\mu \phi \partial^\mu \phi +
	\frac{\lambda}{2} \phi^2 A^a_\mu A^{a \mu} -
	\frac{\Lambda}{4} \left( \phi^2 - M^2 \right)^2.
\label{2_a_10}
\end{equation}
Here
$
	F^a_{\mu \nu} = \partial_\mu A^a_\nu - \partial_\nu A^a_\mu +	g f_{a b c} A^b_\mu A^c_\nu
$ is the field strength tensor for the Proca field, where $f_{a b c}$ are the SU(3) structure constants, $g$ is the coupling constant, $a,b,c = 1,2, \dots, 8$ are color indices, $\mu, \nu = 0, 1, 2, 3$ are spacetime indices. The Lagrangian \eqref{2_a_10} also contains the arbitrary constants $M, \lambda, \Lambda$ and the Proca field ``mass tensor''
$
	\left( \mu^2 \right)^{a b, \mu}_{\phantom{a b,}\nu}
$. We have been using the words ``mass tensor''  in quotation marks because these quantities are not actually masses.
The quantity
 $
\left( \mu^2 \right)^{a b, \mu}_{\phantom{a b,}\nu}
$ corresponds to the case of modified Proca theory where the mass term
$m^2 A^a_\mu A^{a \mu}$ is replaced by
$
\left( \mu^2 \right)^{a b, \mu}_{\phantom{a b,}\nu} A^a_\mu A^{b \nu}
$.
The literature in the field offers a few ways for modifying Proca theory
(see, e.g., Ref.~\cite{BeltranJimenez:2016afo} where generalized multi-Proca theory with the mass term
$M^{ab} A^a_\mu A^{b \mu}$ is under consideration). Following this approach, we have generalized Proca theory by introducing the matrix
 $\left( \mu^2 \right)^{a b, \mu}_{\phantom{a b,}\nu}$ which is a tensor with the spacetime indices  $\mu, \nu$.

Making use of Eq.~\eqref{2_a_10}, the corresponding field equations can be written in the form
\begin{align}
	D_\nu F^{a \mu \nu} - \lambda \phi^2 A^{a \mu} =&
	- \left( \mu^2 \right)^{a b, \mu}_{\phantom{a b,}\nu} A^{b \nu},
\label{2_a_20}\\
	\Box \phi =& \lambda A^a_\mu A^{a \mu} \phi +
	\Lambda \phi \left( M^2 - \phi^2 \right) .
\label{2_a_30}
\end{align}
The required tube is described by the {\it Ans\"{a}tze}
\begin{equation}
	A^2_t = \frac{h(\rho)}{g} , \;
	A^5_z = \frac{v(\rho)}{g} , \;
	A^7_\varphi = \frac{\rho w(\rho)}{g} , \;
	\phi = \phi(\rho),
	\label{2_a_40}
\end{equation}
where $\rho, z,$ and $\varphi$ are cylindrical coordinates. For such a choice of the  SU(3) Proca field potentials, we have the following electric and magnetic field intensities:
\begin{eqnarray}
	&&E^2_\rho = - \frac{h^\prime}{g} , \quad
	E^5_\varphi = - \frac{\rho h w}{2 g} , \quad
	E^7_z = \frac{h v}{2 g} , \nonumber
\label{2_a_50}\\
	&&H^2_\rho = - \frac{v w}{2 g}, \quad
	H^5_\varphi = - \frac{\rho v^\prime}{g}, \quad
	H^7_z = \frac{1}{g} \left(
	w^\prime + \frac{w}{\rho}
	\right) , \nonumber
\label{2_a_60}
\end{eqnarray}
where the prime denotes differentiation with respect to $\rho$. The most interesting fact for us is that the tube contains the longitudinal electric field  $E^7_z \neq 0$ producing the required attraction between quarks described by the Proca field.

Substituting the potentials \eqref{2_a_40} in Eqs.~\eqref{2_a_20} and \eqref{2_a_30}, one can  get the corresponding equations for the functions $h$, $v$, $w$, and $\phi$ (for details see  Ref.~\cite{Dzhunushaliev:2019sxk}).

\begin{figure}[t]
\centering
\includegraphics[width=1\linewidth]{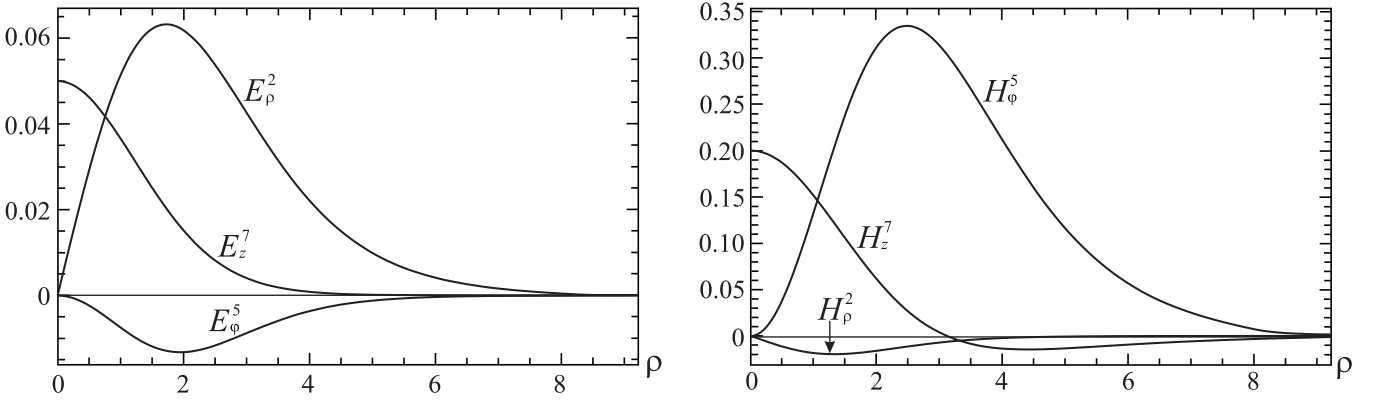}
\vspace{-0.3cm}
\caption{A sketch of the distributions of color electric fields $E^{2}_{\rho}$, $E^{5}_{\varphi}$, $E^{7}_{z}$ and of color magnetic fields $H^{2}_{\rho}$, $H^{5}_{\varphi}$, $H^{7}_{z}$.
}
\label{fig_el_magn_fields}
\end{figure}

A sketch of the distributions of color electric and magnetic fields is given in Fig.~\ref{fig_el_magn_fields}.
As we see, there is the longitudinal electric field $E^7_z$ ensuring the attractive force between quarks and preventing their separation.
Correspondingly, the tube contains a finite flux of this field,
$$
	\Phi_z = 2 \pi \int \limits_0^\infty \rho E^7_z d \rho < \infty ,
$$
whose presence is necessary for the creation of a constant attractive force between quarks.

\subsection{Tube with the energy flux/momentum density
}
\label{tube_2}

\begin{figure}[t]
\centering
\includegraphics[width=1\linewidth]{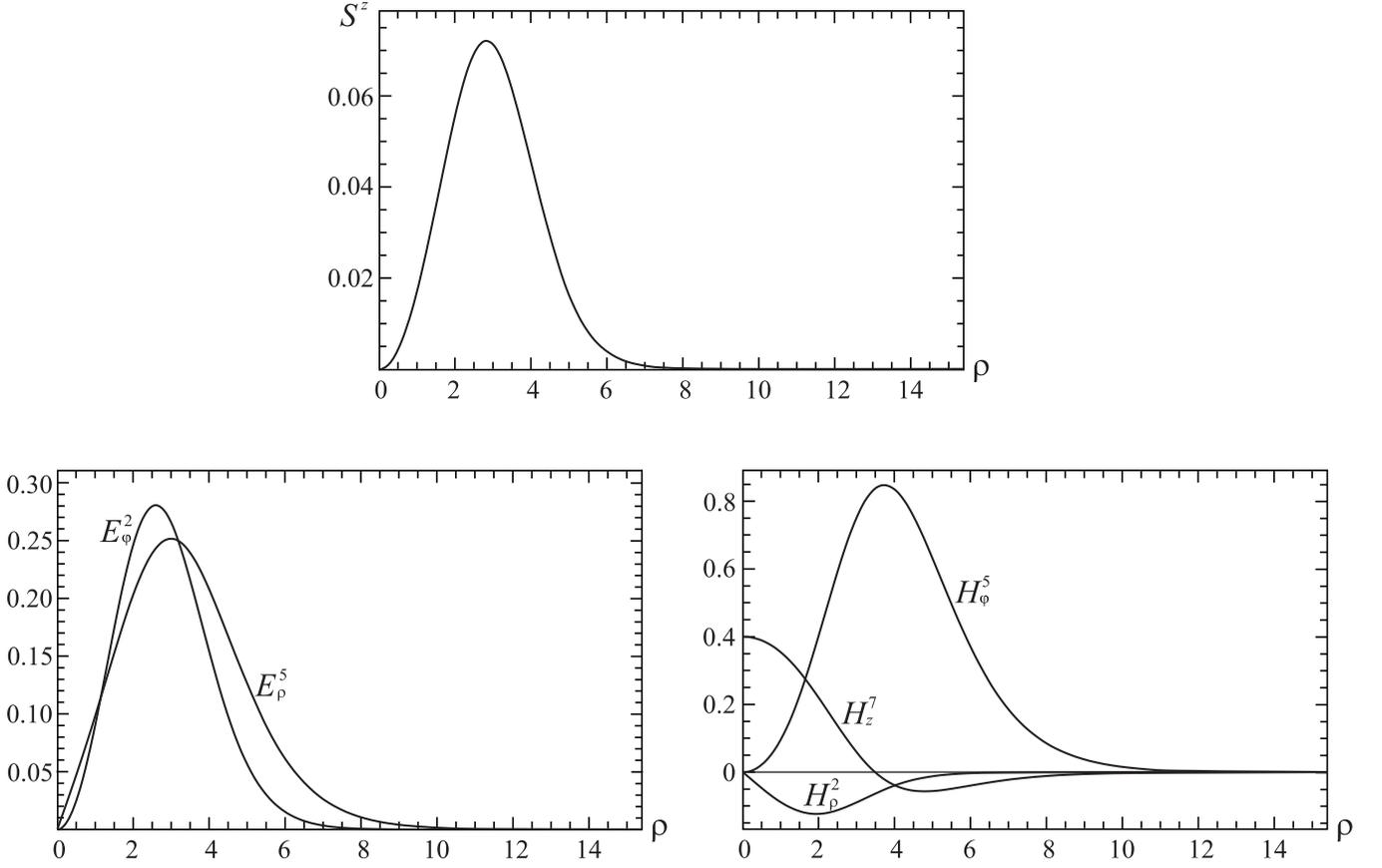}
\vspace{-0.3cm}
\caption{Sketches of the $z$-component of the Poynting vector $S^z$ given by Eq.~\eqref{2_b_90},
of the color electric fields $E^2_\varphi$ and $E^5_\rho$, of the color magnetic fields
$H^2_\rho, H^5_\varphi$, and $H^7_z$.
}
\label{fig_mom_dens}
\end{figure}

In this case we take the {\it Ans\"{a}tze}
\begin{equation}
	A^5_t = \frac{f(\rho)}{g} , \;
	A^5_z = \frac{v(\rho)}{g} , \;
	A^7_\varphi = \frac{\rho w(\rho)}{g} , \;
	\phi = \phi(\rho),
\label{2_b_10}
\end{equation}
which give the following components of the electric and magnetic field intensities:
\begin{align}
	E^2_\varphi =& \frac{\rho f w}{2 g}, \quad
	E^5_\rho = - \frac{f^\prime}{g} , \quad
\label{2_b_20}\\
	H^2_\rho =& - \frac{v w}{2 g}, \quad
	H^5_\varphi = - \frac{\rho v^\prime}{g}, \quad
	H^7_z = \frac{1}{g} \left(
	w^\prime + \frac{w}{\rho}
	\right) .
\label{2_b_30}
\end{align}
In this case the Poynting vector $S^i$ is already nonzero,
\begin{equation}
	S^i = \frac{\epsilon^{i j k}}{\sqrt{\gamma}} E^a_j H^a_k \neq 0,
\label{2_b_40}
\end{equation}
where  $\gamma$ is the determinant of the space metric. Substituting the potentials \eqref{2_b_10} in Eqs.~\eqref{2_a_20} and \eqref{2_a_30}, one can get the corresponding equations for the functions
$f$, $v$, $w$, and $\phi$ (for details see Ref.~\cite{Dzhunushaliev:2019sxk}). In turn,
substitution of the components of electric and magnetic fields \eqref{2_b_20} and \eqref{2_b_30} in Eq.~\eqref{2_b_40} yields the following expression for the Poynting vector:
\begin{equation}
	S^z = \frac{1}{g^2} \left(
		\frac{d f}{d \rho} \frac{d v}{d \rho} + \frac{1}{4}f v w^2
	\right)  .
\label{2_b_90}
\end{equation}
The graph of the momentum density  $S^z$ is schematically sketched in the top panel of Fig.~\ref{fig_mom_dens}.
Its nonzero value is caused by the presence of the crossed electric,  $E^2_\varphi$, and magnetic, $H^2_\rho$, fields,
as well as the electric, $E^5_\rho$, and magnetic, $H^5_\varphi$, fields, whose typical distributions are depicted in the bottom panels of Fig.~\ref{fig_mom_dens}.

\subsection{Proca balls with angular momentum}
\label{Proca_ball}

In this subsection we follow Ref.~\cite{Dzhunushaliev:2021tlm}. The Lagrangian describing a system consisting of a non-Abelian SU(2) Proca field $A^a_\mu$ coupled to a triplet of real Higgs scalar fields $\phi^a$ can be taken in the form
\begin{equation}
	\mathcal L =  - \frac{1}{4} F^a_{\mu \nu} F^{a \mu \nu} -
	\frac{1}{2}\left( \mu^2 \right)^{a b, \mu}_{\phantom{a b,}\nu}
	A^a_\mu A^{b \nu} +
	\frac{1}{2} D_\mu \phi^a D^\mu \phi^a  -
	\frac{\Lambda}{4} \left( \phi^a \phi^a - v^2 \right)^2.
\label{3_c_10}
\end{equation}
Here
$
F^a_{\mu \nu} = \partial_\mu A^a_\nu - \partial_\nu A^a_\mu +
g \epsilon_{a b c} A^b_\mu A^c_\nu
$ is the field strength tensor for the Proca field, where $\epsilon_{a b c}$ are the SU(2) structure constants (the completely antisymmetric Levi-Civita symbol), $g$ is the coupling constant,
$a, b, c = 1, 2, 3$ are SU(2) color indices, $\mu, \nu = 0, 1, 2, 3$ are spacetime indices;
$v$ and $\Lambda$ are arbitrary constants,  and
$
	\left( \mu^2 \right)^{a b, \mu}_{\phantom{a b,}\nu}
$ is the Proca field mass tensor, which we suppose to be symmetric with respect to the color and spacetime indices. This tensor corresponds to the case of modified Proca theory in the same sense as described after Eq.~\eqref{2_a_10}.

Using Eq.~\eqref{3_c_10}, the corresponding field equations can be written in the form
\begin{eqnarray*}
	D_\nu F^{a \mu \nu} +
	\left( \mu^2 \right)^{a b, \mu}_{\phantom{a b,}\nu} A^{b \nu}
	&=& g \epsilon^{abc} \phi^b D^\mu \phi^c ,
\label{3_c_15}\\
	D_\mu D^\mu \phi^a &=&
	- \Lambda \phi^a \left(
	\phi^b \phi^b - v^2
	\right).
\label{3_c_20}
\end{eqnarray*}

We choose the {\it Ans\"{a}tze} for the Proca and scalar fields  in the form
\begin{equation*}
	A^1_t = \frac{ f(\rho, z)}{g} , \;
	A^1_\varphi =  \frac{\rho \, k(\rho, z)}{g} , \;
	A^3_t =  \frac{ h(\rho, z)}{g} , \;
	A^3_\varphi = \frac{\rho \, w(\rho, z)}{g} , \;
	\phi = \left\{\phi_1(\rho, z), 0,  \phi_3(\rho, z)\right\}
	\label{3_c_30}
\end{equation*}
written in cylindrical coordinates  $\{t, \rho, \varphi, z\}$. For such a choice, there are the following
nonvanishing color electric and magnetic fields (physical components):
\begin{align}
	E^1_z = & - \frac{ f_{, z}}{g} , \quad
	E^3_z = - \frac{ h_{, z}}{g}, \quad
	E^1_\rho = - \frac{f_{, \rho}}{g} , \quad
	E^3_\rho = - \frac{h_{, \rho}}{g},
\label{3_c_40}\\
	H^1_z = & - \frac{
		\rho \, k_{, \rho} +  k
	}{g \rho} , \quad
	H^3_z = - \frac{
		\rho \, w_{, \rho} +  w
	}{g\rho} , \quad
	H^1_\rho = \frac{ k_{, z}}{g} ,  \quad
	H^3_\rho = \frac{ w_{, z}}{g} ,
\label{3_c_50}
\end{align}
where a comma in lower indices denotes differentiation with respect to the corresponding coordinate.

The components of the electric field strength given in Eq.~\eqref{3_c_40} can be even or odd functions of
the coordinate $z$. For this reason, we consider here systems of two types:
the cases where the component $E^1_z$ is an odd/even function and  $E^1_\rho$ is an even/odd function. We will refer to such configurations as the type A/B Proca balls (or {\it P}-balls), respectively.

\begin{figure}[H]
	\centering
	\includegraphics[width=0.5\linewidth]{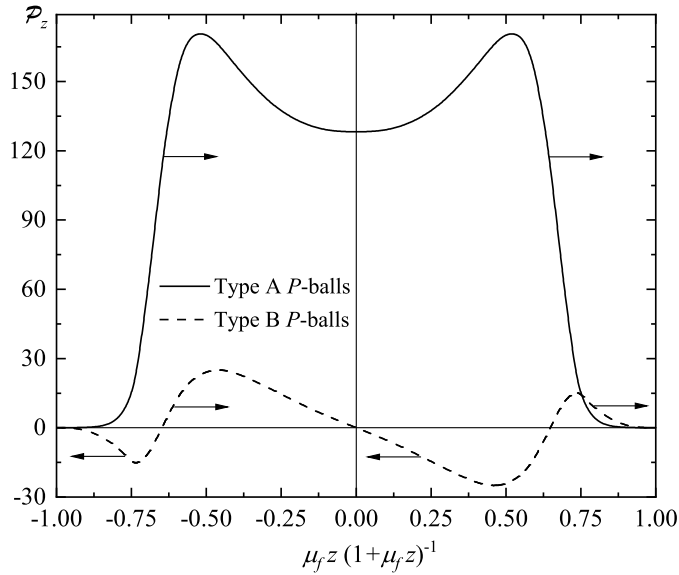}
	\caption{Schematic sketch of the linear angular momentum density ${\mathcal{P}}_z$ defined in Eq.~\eqref{3_c_110} [with the component of the Proca field mass tensor
		$\mu_{f}^2\equiv\left( \mu^2 \right)^{1 1, t}_{\phantom{a b,}t}$] along the tube axis for the {\it P}-balls under consideration. The arrows show the directions of the angular momentum density.	
	}
	\label{P_z_linear}
\end{figure}

For the sake of simplicity, consider the case where
\begin{equation}
	h = - f, k = -w, \phi_1 = \phi_3 = \phi.
\label{3_c_60}
\end{equation}
The corresponding partial differential equations describing such a system can be found in Ref.~\cite{Dzhunushaliev:2021tlm}. In turn, for the field strengths~\eqref{3_c_40} and \eqref{3_c_50}, the expression \eqref{2_b_40} gives the following nonvanishing physical component of the Poynting vector:
\begin{equation*}
	 S_\varphi =
	\frac{2}{g^2}\left[f_{, \rho} \left(w_{, \rho } + \frac{w}{\rho}\right)  +  f_{, z} w_{, z}\right].
\label{3_c_100}
\end{equation*}
Making use of this expression, one can obtain the expression for a linear angular momentum density,
\begin{equation}
	 \mathcal{P}_z =2\pi\int \limits_{0}^{\infty}   S_\varphi \rho^2 d\rho ,
	\label{3_c_110}
\end{equation}
which describes a distribution of the angular momentum density along the tube axis.
The corresponding distributions are exemplified in Fig.~\ref{P_z_linear}.
It is seen from this figure that, for the type A {\it P}-balls, the linear angular momentum is always directed in one direction;
this has the result that such configurations, unlike the type B {\it P}-balls, have a nonzero total angular momentum.

\subsection{Nonperturbative quantization and the hypothetical appearance of a mass term
}
\label{turbulence_modeling}

For the above flux tube and particlelike solutions
to exist, the presence in the system of a massive Yang-Mills field (i.e., a Proca field) is absolutely necessary.
In this subsection we would like to discuss the question of whether, in solving Eqs.~\eqref{NP_10}-\eqref{NP_60} approximately,
there can occur additional terms (in particular, a mass-like term) which might play the role of an effective mass in Proca theory.

The only way to solve the infinite set of equations \eqref{NP_10}-\eqref{NP_60} approximately is the use of the cut-off method for obtaining a finite set of equations. According to this approach, it is necessary to keep only the first $n$ equations. In this case, the last equation will contain higher-order Green's functions ($G_{n+1}$), equations for which are already ruled out from a consideration.  In order that the first $n$ equations form a closed system, it is necessary to propose a hypothesis about the last Green's function $G_{n+1}$. For instance, one can assume that such Green's function is some polylinear combination of lower-order Green's functions. For example, for a fourth-order Green's function, this can be represented as
\begin{equation}
\begin{split}
	& \left\langle
		\hat A^a_\alpha(x_1) \hat A^b_\beta(x_2) \hat A^c_\gamma(x_3) \hat A^d_\delta(x_4)
	\right\rangle \approx
	\left\langle \hat A^a_\alpha(x_1)	\right\rangle \left\langle \hat A^b_\beta(x_2)	\right\rangle
	\left\langle \hat A^c_\gamma(x_3)	\right\rangle \left\langle \hat A^d_\delta(x_4)	\right\rangle
\\
	&
	+ \left( \mu^2 \right)^{a b}_{\phantom{a b,}\alpha \beta}
	\left\langle \hat A^c_\gamma(x_3)	\right\rangle \left\langle \hat A^d_\delta(x_4)	\right\rangle
	+ \left( \text{ transmutation } a,b,c,d; \alpha, \beta, \gamma, \delta
	\right) + C ,
\end{split}
\label{3_d_10}
\end{equation}
where
$
 \hat A^a_\alpha = \left\langle A^a_\alpha \right\rangle + \widehat{\delta A^a_\alpha}
$,
the constants $\left( \mu^2 \right)^{a b}_{\phantom{a b,}\alpha \beta} $ and $C$ are called closure constants.
Here, the lower line is a nonperturbative quantum correction to the first term in the right-hand side of the expression~\eqref{3_d_10}. An assumption of this sort enables one to cut off the infinite set of equations~\eqref{NP_10}-\eqref{NP_60} and to obtain appropriate solutions describing nonperturbative effects in QCD.

We may notice here a remarkable analogy between such procedure of nonperturbative quantization and what happens in turbulence modeling when analysing the Navier-Stokes equation
 \begin{equation*}
	\rho \left(
		\frac{\partial v_i}{\partial t}
	+ v_j  \frac{\partial v_i}{\partial x_j}
	\right) = - \frac{\partial p}{\partial x_i} + \frac{\partial t_{ij}}{\partial x_j},
\label{3_d_20}
\end{equation*}
where $v_i$ is the flow velocity, $\rho$ is the fluid density, $p$ is the pressure, $t_{ij} = 2 \mu s_{ij}$ is the viscous stress tensor,
$s_{ij} = \frac{1}{2} \left(
\frac{\partial v_i}{\partial x_j} + \frac{\partial v_j}{\partial x_i}
\right)$, $\mu$ is molecular viscosity, and the velocity $\vec v$ is a fluctuating quantity. Averaging this equation, one can obtain~\cite{Wilcox}
\begin{equation}
		\rho \frac{\partial V_i}{\partial t} +
	\rho V_j \frac{\partial V_i}{\partial x_j}
	= - \frac{\partial p}{\partial x_i} +
	\frac{\partial }{\partial x_j} \left(
	2 \mu S_{ji} - \overline{\rho v_j^\prime v_i^\prime}
	\right),
\label{3_d_40}
\end{equation}
where the mean velocity $V_i$ is defined according to the equation
$v_i(\vec x, t) = V_i(\vec x, t) + v^\prime_i(\vec x, t)$. Eq.~\eqref{3_d_40} is usually referred to as
the Reynolds-averaged Navier-Stokes equation. The quantity $\overline{\rho v_j^\prime v_i^\prime}$ is known as the Reynolds-stress tensor and denoted as
\begin{equation*}
	\tau_{ij} = - \overline{\rho v_i^\prime v_j^\prime}.
\label{3_d_50}
\end{equation*}
One can obtain the following equation for the Reynolds stress tensor:
\begin{equation*}
	\begin{split}
		\frac{\partial \tau_{ij}}{\partial t} +
		V_k \frac{\partial \tau_{ij}}{\partial x_k} =
		- \tau_{ik} \frac{\partial V_j}{\partial x_k} -
		\tau_{jk} \frac{\partial V_i}{\partial x_k} +
		2 \mu \overline{\frac{\partial v^\prime_i}{\partial x_k}
			\frac{\partial v^\prime_j}{\partial x_k}} -
		\overline{
			p^\prime \left(
			\frac{\partial v_i^\prime}{\partial x_j} +
			\frac{\partial v_j^\prime}{\partial x_i}
			\right)
		} +
		\frac{\partial}{\partial x_k} \left(
		\nu \frac{\partial \tau_{ij}}{\partial x_k} +
		C_{ijk}
		\right),
\label{3_d_60}
\end{split}
\end{equation*}
where $\nu$ is kinematic viscosity and
\begin{equation*}
	C_{ijk} =
	\overline{\rho v^\prime_i \rho v^\prime_j \rho v^\prime_k} +
	\delta_{jk} \overline{p^\prime v^\prime_i} +
	\delta_{ik} \overline{p^\prime v^\prime_j}.
\end{equation*}
Thus we have six new equations, one for each independent component of the Reynolds-stress tensor $\tau_{ij}$. However, we have also generated new unknowns
$
	\overline{\rho v^\prime_i \rho v^\prime_j \rho v^\prime_k}$,
$\overline{\mu \frac{\partial v^\prime_i}{\partial x_k}
	\frac{\partial v^\prime_j}{\partial x_k}}$,
$\overline{v^\prime_i \frac{\partial p^\prime}{\partial x_j}}$, and
$\overline{v^\prime_j \frac{\partial p^\prime}{\partial x_i}}$.
In the same way one can obtain equations for these new unknown cumulants, and so on, {\it ad infinitum}.
In order to cut off this infinite set of equations, one uses the procedure known as the closure problem in turbulence modeling.

Thus, returning to the closure equation~\eqref{3_d_10}, we see that the approximate approach to solving the infinite set of equations of nonperturbative quantization \eqref{NP_10}-\eqref{NP_60} leads to the appearance of the closure constants, one of which may be a mass of a Yang-Mills field. It is quite conceivable that in the limit $n \rightarrow \infty$ some of such constants will remain. This means that the nonperturbative quantization leads to the appearance of dimensional constants which are absent in the initial classical theory; this phenomenon is called dimensional transmutation~\cite{Coleman:1973jx}.

\section{Discussion and conclusions}

In this study, we have shown that some QCD effects can occur in non-QCD theories which differ from QCD by the presence, for example, a nonlinear spinor field or a massive Yang-Mills field (a Proca field). For this purpose, we have discussed hypothetical possibilities of the appearance of such effects in QCD as a consequence of quantization of essentially nonlinear  Yang-Mills fields. We have found out that when one uses the approximate approach to solving an infinite set of Dyson-Schwinger equations describing the procedure of nonperturbative quantization, the question arises as to how one may describe a higher-order Green's function in the cut-off set of equations.

In the present paper we have considered two possible ways of solving this problem related to
(i)~the use of the nonlinearity describing the interaction of gauge and spinor fields and (ii)~the nonlinearity of the gauge field itself:
\begin{itemize}
	\item Green's function describing the interaction between gauge and spinor fields (for instance,
	$\left\langle \hat{\bar \psi} \gamma^\mu \hat{A}^a_\mu \hat \psi \right\rangle$) is approximately described as
the sum of two terms. The first term breaks up the quantum average of the product of two fields as the product
of the averages of these fields, i.e., it is the product of the two-point Green's function of the spinor field
	$
		\left\langle \bar{\hat \psi} \hat \psi \right\rangle^2
	$
by the average of the gauge field $\left\langle \hat A^a_\mu \right\rangle$.
The second term is the correction, and it consists of the product of two factors: the first one is the square of the two-point Green's function
	$
		\left\langle \bar{\hat \psi} \hat \psi \right\rangle^2
	$, and the second one is the two-point Green's function of the Yang-Mills fields
	$
		\left\langle \widehat{\delta A^a_\mu} \widehat{\delta A^{a \mu}} \right\rangle
	$, i.e., it is the dispersion of this field.
	\item Green's function of the product of Yang-Mills field potentials is a combination of lower-order Green's functions of the gauge field,
and the so-called closure constants appear, among which there may be masses of the gauge field and $\Lambda_{\text{QCD}}$ (a characteristic parameter coming from QCD).
\end{itemize}

As concrete examples, we have demonstrated that there exist the following QCD effects in non-QCD theories:
\begin{itemize}
	\item In SU(2) Yang-Mills theory with a source in the form of a nonlinear spinor field, a mass gap is present.
	\item In non-Abelian Proca theories, there are:
		\begin{itemize}
			\item Infinite tubes containing a longitudinal electric field (see Sec.~\ref{tube_1}).
			\item Infinite tubes with the momentum density created by crossed electric and magnetic fields and directed along the tube axis (see Sec.~\ref{tube_2}).
			\item Particlelike solutions possessing a nonzero total angular momentum created by crossed electric and magnetic fields  (see Sec.~\ref{Proca_ball}).
			\item For all the above solutions, the Meissner-like effect is observed: the Proca fields are expelled by the Higgs fields.
		\end{itemize}
\end{itemize}

In connection with the conjecture made here, there arise the following questions requiring more detailed investigations:
\begin{itemize}
	\item What happens with the closure constants in the limit $n \rightarrow \infty$? If the closure constants survive in this limit,
this may lead to the appearance of dimensional constants which are absent in the initial classical theory.
This means that there will take place dimensional transmutation for nonperturbative quantization.
	\item Whether there exists any generalization of the procedure of renormalization for the procedure of  nonperturbative
quantization described in Sec.~\ref{NP_quantization}? 	\item How deep is the analogy between nonperturbative quantization and turbulence modeling pointed out in Sec.~\ref{turbulence_modeling}?
\end{itemize}

Thus the main result of the present study is that we make the conjecture that the approximate approach to the procedure of quantization of strongly interacting fields (the procedure of  nonperturbative quantization) may lead to the appearance of either new nonlinearities or new terms in the corresponding equations.

\section*{Acknowledgments}

The work was supported by the program No.~AP14869140  (The study of QCD effects in non-QCD theories) of the Ministry of Education and Science of the Republic of Kazakhstan. We are also grateful to the Research Group Linkage Programme of the Alexander von Humboldt Foundation for the support of this research. We wish to thank the anonymous referees whose comments helped to improve the manuscript.


\begin{thebibliography}{99}

\bibitem{tinkham}
M.~Tinkham, {\it  Introduction to Superconductivity} (McGraw-Hill,  1996).

\bibitem{Klevansky:1992qe}
S.~P.~Klevansky,
The Nambu-Jona-Lasinio model of quantum chromodynamics,
Rev. Mod. Phys. \textbf{64}, 649 (1992).

\bibitem{Greensite:2011zz}
J.~Greensite, An introduction to the confinement problem,
Lect. Notes Phys. {\bf 821}, 1 (2011).
	
\bibitem{Hasenfratz:1977dt}
P.~Hasenfratz and J.~Kuti,
The Quark Bag Model,
Phys. Rept. \textbf{40}, 75 (1978).

\bibitem{Ashman:1987hv}
J.~Ashman {\it et al.} [European Muon Collaboration],
A Measurement of the Spin Asymmetry and Determination of the Structure Function g(1) in Deep Inelastic Muon-Proton Scattering,
Phys.\ Lett.\ B {\bf 206}, 364 (1988) .
	
\bibitem{Ashman:1989ig}
J.~Ashman {\it et al.} [European Muon Collaboration],
An Investigation of the Spin Structure of the Proton in Deep Inelastic Scattering of Polarized Muons on Polarized Protons,
Nucl.\ Phys.\ B {\bf 328}, 1 (1989).

\bibitem{Ratti:2018ksb}
C.~Ratti,
Lattice QCD and heavy ion collisions: a review of recent progress,
Rept. Prog. Phys. \textbf{81},  084301 (2018).

\bibitem{Shnir:2005vvi}
Ya.~Shnir, {\it Magnetic Monopoles} (Springer, Berlin-Heidelberg-New York, 2005).

\bibitem{Shuryak:2021vnj}
E.~Shuryak,
Nonperturbative Topological Phenomena in QCD and Related Theories,
Lect. Notes Phys. \textbf{977}, 1 (2021).
	
\bibitem{Dzhunushaliev:2020qwf}
V.~Dzhunushaliev, V.~Folomeev, and A.~Serikbolova,
Monopole solutions in SU(2) Yang-Mills-plus-massive-nonlinear-spinor-field theory,
Phys.\ Lett.\ B {\bf 806}, 135480 (2020).

\bibitem{Dzhunushaliev:2021apa}
V.~Dzhunushaliev, N.~Burtebayev, V.~N.~Folomeev, J.~Kunz, A.~Serikbolova, and A.~Tlemisov,
Mass gap for a monopole interacting with a nonlinear spinor field,
Phys. Rev. D \textbf{104}, 056010 (2021).

\bibitem{heis}
W. Heisenberg, \textit{Introduction to the unified field theory of elementary particles} (Max-Planck-Institut f\"ur Physik und Astrophysik, Interscience Publishers London, New York, Sydney, 1966).

\bibitem{Bender:1999ek}
C.~M.~Bender, K.~A.~Milton, and V.~Savage,
Solution of Schwinger-Dyson equations for PT symmetric quantum field theory,
Phys. Rev. D \textbf{62}, 085001 (2000).

\bibitem{Frasca:2019ysi}
M.~Frasca,
Differential Dyson\textendash{}Schwinger equations for quantum chromodynamics,
Eur. Phys. J. C \textbf{80},  707 (2020).

\bibitem{Coleman:1973jx}
S.~R.~Coleman and E.~J.~Weinberg,
Radiative Corrections as the Origin of Spontaneous Symmetry Breaking,
Phys. Rev. D \textbf{7}, 1888 (1973).

\bibitem{Dzhunushaliev:2019sxk}
V.~Dzhunushaliev and V.~Folomeev,
Proca tubes with the flux of the longitudinal chromoelectric field and the energy flux/momentum density,
Eur. Phys. J. C \textbf{80}, 1043 (2020).

\bibitem{BeltranJimenez:2016afo}
J.~Beltran Jimenez and L.~Heisenberg,
Generalized multi-Proca fields,
Phys. Lett. B \textbf{770}, 16 (2017). 

\bibitem{Dzhunushaliev:2021tlm}
V.~Dzhunushaliev and V.~Folomeev,
Proca balls with angular momentum or flux of electric field,
Phys. Rev. D \textbf{105}, 016022 (2022).

\bibitem{Wilcox}
D.~C.~Wilcox,
{\it Turbulence Modeling for CFD} (DCW Industries, Inc.
La Canada, California, 1994).

\end{thebibliography}
\end{document}